\documentclass[preprint]{aastex}
\usepackage{natbib}
\usepackage{color}
\usepackage{ulem}
\slugcomment{2009, ApJ, 700L, 127A}

\shorttitle{Milagro Observations of Multi-TeV Emission from Galactic Sources in the Fermi-LAT Bright Source List}
\shortauthors{Abdo et al.}

\begin{document}


\title{Milagro Observations of Multi-TeV Emission from Galactic Sources in the Fermi Bright Source List}

\author{
A.~A.~Abdo,\altaffilmark{\ref{msu},\ref{nrl}}
B.~T.~Allen,\altaffilmark{\ref{uci},\ref{cfa}} 
T.~Aune,\altaffilmark{\ref{ucsc}}
D.~Berley,\altaffilmark{\ref{umcp}} 
C.~Chen,\altaffilmark{\ref{uci}}
G.~E.~Christopher,\altaffilmark{\ref{nyu}}
T.~DeYoung,\altaffilmark{\ref{psu}} 
B.~L.~Dingus,\altaffilmark{\ref{lanl}} 
R.~W.~Ellsworth,\altaffilmark{\ref{georgemason}} 
M.~M.~Gonzalez,\altaffilmark{\ref{ida}} 
J.~A.~Goodman,\altaffilmark{\ref{umcp}} 
E.~Hays,\altaffilmark{\ref{gsfc}},
C.~M.~Hoffman,\altaffilmark{\ref{lanl}}
P.~H.~H\"untemeyer,\altaffilmark{\ref{utah}}
B.~E.~Kolterman,\altaffilmark{\ref{nyu}} 
J.~T.~Linnemann,\altaffilmark{\ref{msu}}
J.~E.~McEnery,\altaffilmark{\ref{gsfc}}
T.~Morgan,\altaffilmark{\ref{unh}}
A.~I.~Mincer,\altaffilmark{\ref{nyu}} 
P.~Nemethy,\altaffilmark{\ref{nyu}} 
J.~Pretz,\altaffilmark{\ref{lanl}}
J.~M.~Ryan,\altaffilmark{\ref{unh}} 
P.~M.~Saz~Parkinson,\altaffilmark{\ref{ucsc}}
A.~Shoup,\altaffilmark{\ref{osu}} 
G.~Sinnis,\altaffilmark{\ref{lanl}} 
A.~J.~Smith,\altaffilmark{\ref{umcp}} 
V.~Vasileiou,\altaffilmark{\ref{umcp},\ref{cresst}} 
G.~P.~Walker,\altaffilmark{\ref{lanl},\ref{nst}} 
D.~A.~Williams\altaffilmark{\ref{ucsc}}
and 
G.~B.~Yodh\altaffilmark{\ref{uci}}} 

\altaffiltext{1}{\label{msu} Department of Physics and Astronomy, Michigan State University, 3245 BioMedical Physical Sciences Building, East Lansing, MI 48824}
\altaffiltext{2}{\label{nrl} Current address: Space Science Division, Naval Research Laboratory, Washington, DC 20375}
\altaffiltext{3}{\label{uci} Department of Physics and Astronomy, University of California, Irvine, CA 92697}
\altaffiltext{4}{\label{cfa} Current address: Harvard-Smithsonian Center for Astrophysics, Cambridge, MA 02138}
\altaffiltext{5}{\label{ucsc} Santa Cruz Institute for Particle Physics, University of California, 1156 High Street, Santa Cruz, CA 95064}
\altaffiltext{6}{\label{umcp} Department of Physics, University of Maryland, College Park, MD 20742}
\altaffiltext{7}{\label{nyu} Department of Physics, New York University, 4 Washington Place, New York, NY 10003}
\altaffiltext{8}{\label{psu} Department of Physics, Pennsylvania State University, University Park, PA 16802}
\altaffiltext{9}{\label{lanl} Group P-23, Los Alamos National Laboratory, P.O. Box 1663, Los Alamos, NM 87545}
\altaffiltext{10}{\label{georgemason} Department of Physics and Astronomy, George Mason University, 4400 University Drive, Fairfax, VA 22030}
\altaffiltext{11}{\label{ida} Instituto de Astronom\'ia, Universidad Nacional Aut\'onoma de M\'exico,
D.F., M\'exico, 04510}
\altaffiltext{12}{\label{gsfc} NASA Goddard Space Flight Center, Greenbelt, MD 20771}
\altaffiltext{13}{\label{utah} Department of Physics, University of Utah, Salt Lake City, UT 84112}
\altaffiltext{14}{\label{unh} Department of Physics, University of New Hampshire, Morse Hall, Durham, NH 03824} 
\altaffiltext{15}{\label{osu} Ohio State University, Lima, OH 45804}
\altaffiltext{16}{\label{cresst} CRESST NASA/Goddard Space Flight Center, MD 20771 and University of Maryland, Baltimore County, MD 21250}
\altaffiltext{17}{\label{nst} Current address: National Security Technologies, Las Vegas, NV 89102}

\begin{abstract}

We present the result of a search of the Milagro sky map for spatial
correlations with 
sources from a subset of the recent Fermi Bright Source List (BSL).
The BSL consists of the 205 most significant sources detected above 100 MeV by the
Fermi Large Area Telescope.
We select sources based on their categorization in the 
BSL, taking all confirmed or possible Galactic sources
in the field of view of Milagro.  Of the
34 Fermi sources selected, 14 are observed by Milagro at a significance
of 3 standard deviations or more.
We conduct this search with a new analysis which employs
newly-optimized gamma-hadron separation 
and utilizes the full 8-year Milagro dataset.
Milagro is sensitive to gamma rays with energy from 1 to 100 TeV with a peak 
sensitivity from 10-50 TeV depending on the source spectrum and declination. 
These results
extend the observation of these sources far above the Fermi energy band.
With the new analysis and additional data, multi-TeV emission is definitively observed
associated with the Fermi pulsar, J2229.0+6114, in the
Boomerang Pulsar Wind Nebula (PWN).
Furthermore, an extended region of multi-TeV 
emission is associated with the 
Fermi pulsar, J0634.0+1745,
the Geminga pulsar.

\end{abstract}



\section{Introduction}

The Milagro gamma-ray observatory has performed the most sensitive survey of 
1 to 100 TeV gamma
rays from the Northern Hemisphere sky \citep{milagrocygnus,milagrogalacticplane}. 
The Milagro data set is ideal for 
searching for new classes of  
gamma-ray sources.  The recent release of the Bright 
Source List (BSL) by the Fermi collaboration \citep{fermibsl}
 presents such an opportunity by 
looking for coincidences of $>1$ TeV emission with these GeV sources.      
There are 34 sources in the BSL within Milagro's field of view that are not associated 
with extragalactic sources.  
We present 
a search of the Milagro data for excesses between 1 and 100 TeV 
coincident with these 34 
potential Galactic sources.
The analysis presented here differs from previous analyses
\citep{milagrocygnus,milagrogalacticplane,milagrodiffuse} by
optimizing 
the event weighting and Gaussian weighting separately in bins of
event size (measured with the fraction of channels hit in the instrument).
With the improved analysis and an additional year and a half
of data,
the sensitivity has increased by 15\% to
25\%, depending on the spectrum of the source.  


\section{Analysis and Results}

We select Fermi-LAT sources in the field of view of Milagro (with $\delta >$ -5$^\circ$)
based on their categorization in the BSL.  Sources are selected
which are 
confirmed or potential Galactic sources.  Sources that are identified as extragalactic
are omitted.
Sixteen of the selected sources were categorized in the BSL as confirmed pulsars (PSR) and one is a 
high-mass X-ray binary (HXB).
Five sources have a potential association with an SNR,
and 12 have no clear association.
For each of these 34 sources, we calculate
the statistical significance of the Milagro data at the 
BSL position and
estimate the flux or flux limit 
under the assumption that the 
emission is from a single point source.

The flux measurements given in Table \ref{results} are derived with a
similar approach to \citep{milagrogalacticplane}.  The flux is measured 
with an assumed spectrum of $E^{-2.6}$ without a cutoff.  
The dependence of the calculated flux on the true spectrum
is minimized when the flux is quoted at the median energy of the hypothesized 
spectrum.  The median energy depends on Declination and varies between 
32 and 46 TeV for $\delta$ in the range of 10$^\circ$ and 60$^\circ$.  
At a Declination of -5$^\circ$, the median energy of the hypothesized spectrum 
is 90 TeV.  We quote the flux for all sources 
above 3$\sigma$ at a representative value of 35 TeV. It should be noted 
that the median energy used is for the assumed spectrum and not 
experimentally measured. In particular, a source may in fact cut off 
before 35 TeV (the Crab for example) and our analysis 
would still report a flux at 35 TeV. The energy spectrum 
of each reported source is the subject of a paper in preparation.


The results of this search are summarized in Table \ref{results}.
Of the 34 targets,
14 have a significance greater than 3$\sigma$.  
Six of these are associated
with sources or candidates from the first Milagro survey of the Galactic plane 
\citep{milagrogalacticplane}.  
The Crab, MGRO 2019+37, MGRO 1908+06, MGRO 2031+41, and Milagro 
candidates C3 (likely associated with Geminga) and 
C4 (likely associated
with the Boomerang PWN) are all near LAT GeV sources.
In the Milagro data set, 
the 3$\sigma$-5$\sigma$ observations are fairly marginal because they
cannot be convincingly 
discerned from background when statistical penalties for searching 
the
entire sky are taken into account.  
However, with LAT points as a trigger
for the search, the statistical penalties are reduced.  The probability
of a single 3$\sigma$ false-positive in 34 samples of pure background
is only $\sim$4.4\%. 
The probability of 4 or more excesses at or 
above 3$\sigma$ in 34 trials is
$\sim1.5 \times 10^{-7}$.
It is very likely that most of 
our 3$\sigma$ excesses are due to multi-TeV emission\footnote{
Alternatively, using the False Discover Rate method \citep{fdr1,fdr2} 
and requiring an 
estimate of 1\% of the members of the selected candidates to be a 
false discovery, gives the same list of candidates.   
Changing the contamination fraction criterion from .01 to .001 (or to 0.1), 
would have included one fewer (or 3 more) sources, respectively.
}.  
We, therefore, see strong evidence for multi-TeV emission associated with 
Galactic LAT BSL sources
as a class, even if individual sources are not strong enough to definitively
distinguish.

There is some contribution to these measurements from the Galactic diffuse
emission, but that contribution is small.
We can make a conservative estimate by taking the 
Milagro measurement 
of the diffuse emission \citep{milagrodiffuse} at its highest value,
in the inner Galaxy ($30^{\circ} < l < 65^{\circ}$, $|b| < 2^{\circ}$). 
Using this value, we
expect $5.3 \times 10^{-17}$ $\rm{TeV}^{-1} \rm{s}^{-1} \rm{cm}^{-2}$ 
in a 1$^\circ$ bin 
at 35 TeV, 
which is only about $\sim$15\% contamination for the weakest sources in 
Table \ref{sources}.  
The GALPROP conventional model, for comparison, would only
constitute $\sim$3\% contamination.
The contamination is likely lower than suggested by the Milagro 
measurement because of unresolved 
sources, such as many of the sources from Table \ref{results}.
It has even been suggested \citep{casanovadingusdiffuse} that
most of the Milagro diffuse measurement could be due to unresolved
sources.
Finally, the Fermi points observed at 3$\sigma$ in the Milagro 
data occur 
near local maxima in the Milagro data.  
In contrast, the diffuse emission is expected to vary slowly across the Galaxy.

\section{Discussion}

From this analysis, it appears quite common for Galactic 100 MeV - 100 GeV 
sources to have 
associated multi-TeV emission.  This association is notable for pulsars, 
where 9 of 16 
pulsars from the BSL are on our list of likely multi-TeV emitters.  
The pulsars in the BSL which have less than 3$\sigma$ significance 
in Milagro data tend to lie off the Galactic 
plane.  
The pulsars off the plane
are typically older, having traveled far from their origin after the kick 
they received from the
initial asymmetric supernova \citep{pulsarkick}.
Of the SNR sources on the list, 
we see 3 of 5.  Interestingly, 
we see only 2
of the 12 unidentified sources.  These unidentified sources may
be extragalactic and not visible with this analysis which was
optimized for high-energy emission.

\begin{figure}
\includegraphics[width=180mm]{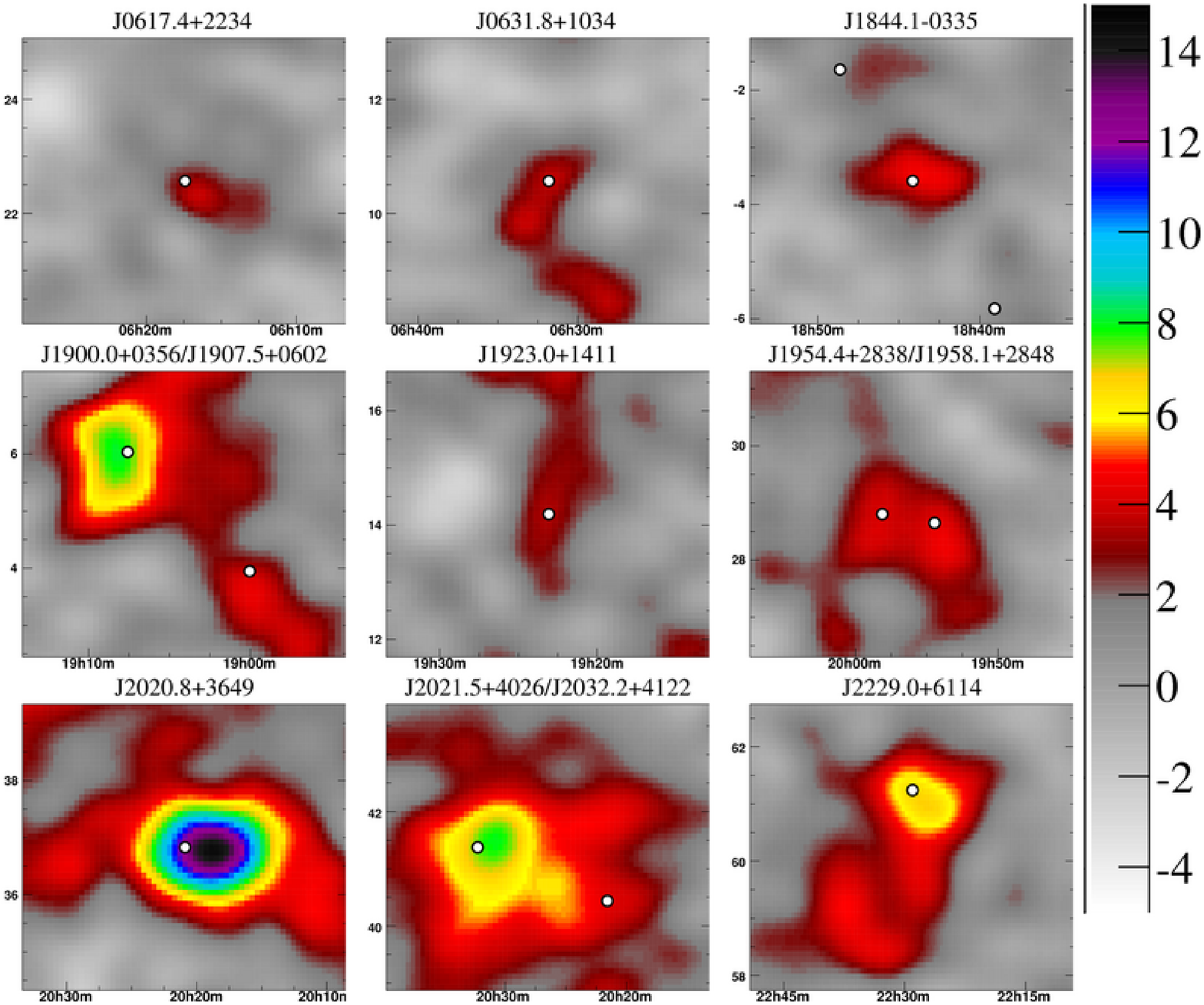}
\caption{\label{sources}
The $3\sigma$ sources from Table \ref{results}, omitting
J0634.0+1745 (shown in Figure \ref{geminga}) and the Crab.  Each frame shows
a 5$^\circ$x5$^\circ$ region with the LAT source
positions 
indicated by 
white dots.  The error on the Fermi source locations
is quite small on this scale, typically between 0.1 and 0.2 degrees, depending 
on the source.  The data has been
smoothed by a Gaussian of width varying between 0.4$^\circ$ and
1.0$^\circ$, depending on the expected angular resolution
of events.
Horizontal axes show
Right-Ascension and vertical axes show
Declination.  
The colors indicate the statistical
significance in standard deviations.
}
\end{figure}

\begin{figure}
\plottwo{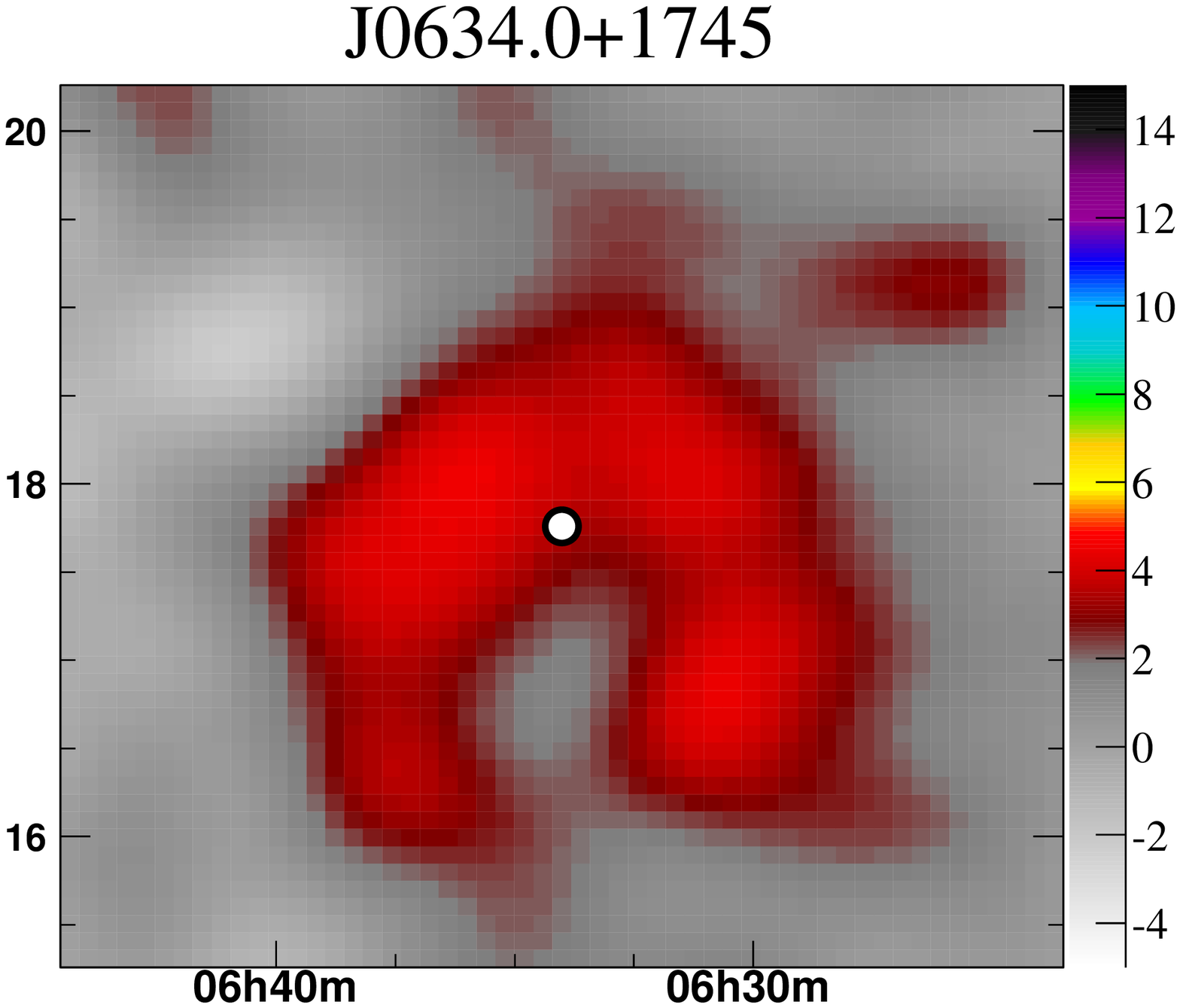}{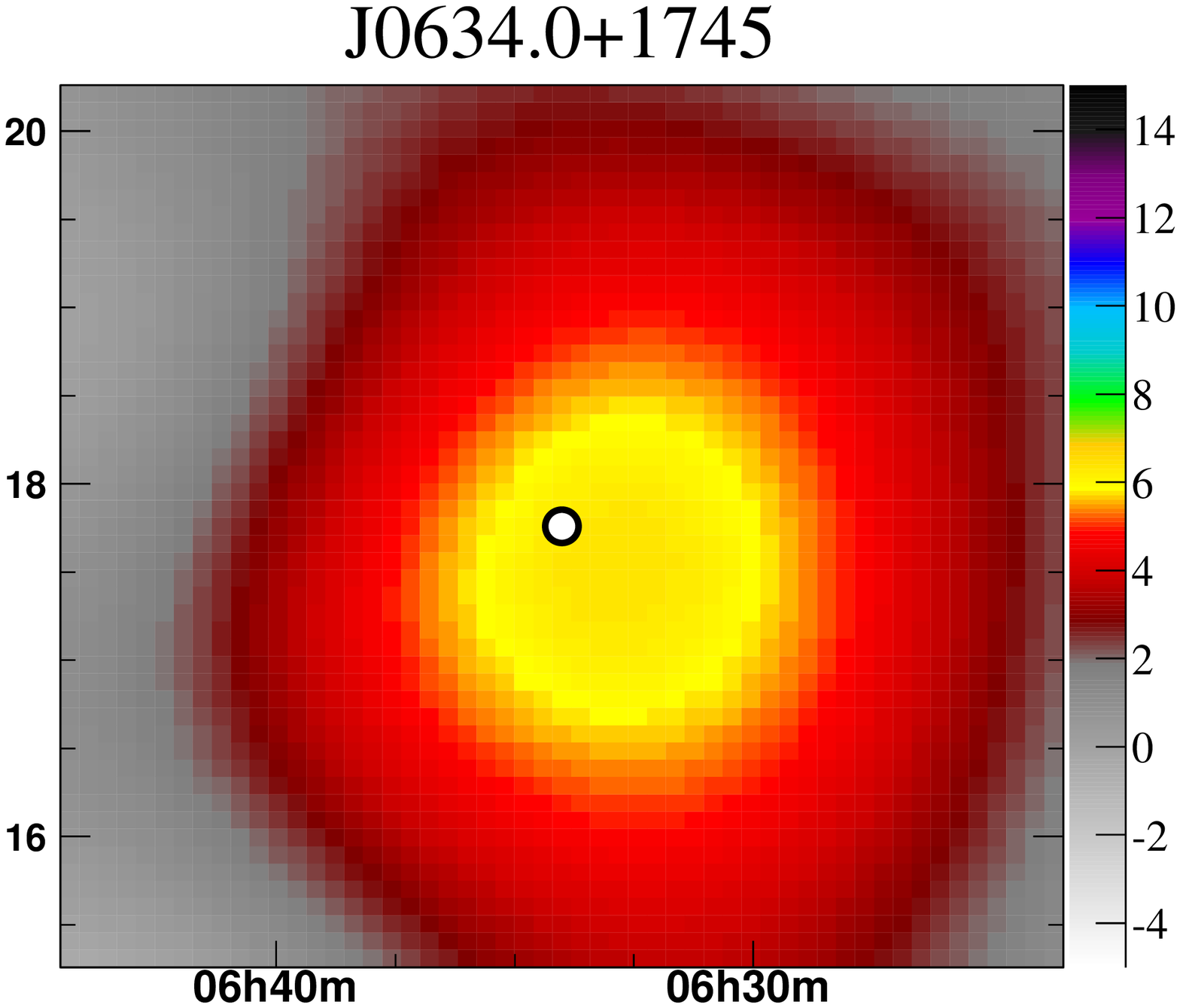}
\caption{\label{geminga}
The signifcance of the Milagro data in a 
5$^\circ$x5$^\circ$ region
around Fermi source J0634.0+1745
the Geminga pulsar.  The location of the Fermi source
is identified by a white dot.  
The figure on the left shows the significance 
map after smoothing by the Milagro point-spread function.  The
figure on the right shows the same region smoothed by an additional
1$^\circ$ Gaussian in order to search for an extended emission region.  
The color scale shows the statistical significance.
}
\end{figure}

Figure \ref{sources} and \ref{geminga} 
shows the regions in the Milagro data around the 
indicated LAT sources. Eight of the 13 sources are associated with
previously reported $>$TeV sources or candidates:


{\bf 0FGL J0534.6+2201} is the young Crab Pulsar.
Its associated pulsar wind nebula (PWN)
is a standard reference source in TeV astronomy.  

{\bf 0FGL J0617.4+2234} is associated with SNR IC443, which is
interacting with a nearby large molecular cloud.  
An associated x-ray feature
has been interpreted as a PWN \citep{ic443xray}, implying the
existence of a pulsar, but the no pulsed emission has yet been detected.
IC443 was first reported above 1 TeV 
by MAGIC \citep{magicic443} and later confirmed by VERITAS \citep{ic443veritas}.  
The flux reported in Table \ref{results}
is somewhat higher than the flux predicted by extrapolating the MAGIC fit, 
but is
roughly consistent after allowing for the extremes of the statistical and 
systematic errors of the two measurements.

{\bf 0FGL J0634.0+1745} is the Geminga pulsar.
Geminga is a relatively old (342 kyr) 
but very near (169 pc) pulsar \citep{atnfcatalog,gemingapulsar}.  It is the 
most significant Fermi-LAT 
source in the northern sky, but emission over 1 TeV
has only been reported 
by Milagro as candidate C3 with too low a significance to be classified as 
a definitive detection. 
Milagro observes an emission region that is 
extended by several degrees as shown in Figure \ref{geminga}. 
The significance reported in Table 
\ref{results} has been computed assuming point source emission, but
if we instead assume that the source is due to emission from an extended 
region and convolve a 1$^{\circ}$ Gaussian with the energy-dependent point 
spread function, the significance at the location of 0FGL J0634.0+1745 
increases to 6.3$\sigma$. 
The local maximum of the Milagro excess is at RA=6h32m28s, Dec=17$^\circ$22m.
Given the high significance, we regard this as a definitive detection of
extended emission from Geminga. 
A spatial Gaussian fit to the data yields a region with a standard 
deviation of 
1.30$^\circ$$\pm0.20^\circ$.  
For comparison, the analogous fit for the Crab, which is effectively
a point source, has a $\sigma$ of 0.6$^\circ$
This
suggests that the full width at half maximum 
of the region of emission in the vicinity of Geminga
is ${2.6^{+0.7}_{-0.9}}^\circ$, after accounting for the point spread function.  
The large extent (implying an emission region of some 5 to 10 pc extent)
is likely due to the 
nearness of the source and may arise from a pulsar-driven 
wind; 
it is consistent with HESS observation of more distant PWN with an 
angular size of $\sim$10 pc.  This may also explain why the source has not
yet been observed by 
Imaging Atmospheric Cherenkov Telescopes \citep{veritasgeminga}.

{\bf 0FGL J1907.5+0602} is associated with MGRO J1908+06\citep{milagrogalacticplane}. 
This pulsar was discovered by the LAT and is also coincident with 
AGILE source
1AGL J1908+0613 \citep{agilecatalog}
and EGRET source GEV J1907+557 \citep{egretgev}.
The multi-TeV emission was first reported by Milagro.
HESS both confirmed the Milagro detection and was also able
to identify this source as extended by $0.21{^{+0.07}_{-0.05}}^\circ$ \citep{hess1908}. 
The peak of the Milagro detection occurs
at RA=19h6m44s, Dec=5$^\circ$50m  with a 1 sigma error circle of 0.27$^\circ$ 
and a local peak significance of 8.1$\sigma$.
The peak of the Milagro emission is 0.3$^\circ$ from the pulsar, but consistent 
with the pulsar's location within the measurement error.

{\bf 0FGL J1923.0+1411} is associated with 
SNR G49.2-0.7 (W51) which is in a star-forming region and near
molecular clouds.  Recently, a $>$TeV source, HESS J1923+141 \citep{hessj1923+141},
has been detected which is spatially extended and coincident 
with the Fermi source.

{\bf 0FGL J2020.8+3649} is associated with MGRO J2019+37.  
This is the most significant source in the Milagro data set apart from the Crab. 
The young central pulsar has a period of 104 ms and an estimated age of 17.2 kyr. 
This source was also detected by AGILE and EGRET.  It was AGILE that 
first identified the GeV pulsations \citep{agile2021plus3651} and 
that discovery was confirmed with Fermi data.  The peak of 
the flux 
measured by Milagro is at RA=20h18m43s Dec=36$^\circ$42m with a 0.09$^\circ$ 1-sigma 
error circle.
The position of the excess is $\sim$0.3$^\circ$ from the pulsar.

{\bf 0FGL J2032.2+4122} is a LAT identified pulsar that is spatially coincident with 
the HEGRA source J2032+41 \citep{hegra2032}, MGRO J2031+41, and the
MAGIC source J2032+4130 \citep{magic2032+4130}.  The Milagro source was reported 
\citep{milagrogalacticplane} with an extent of 
3$^{\circ}$, but it appears that the Milagro extended source may be due to two or more 
overlapping sources with a potential additional diffuse contribution from the highly
emissive Cygnus region.  
The location of the Milagro peak is RA=20h31m43s and Dec=40$^\circ$40m with
a statistical error of 0.3$^\circ$.

{\bf 0FGL J2229.0+6114} is coincident with the radio pulsar J2229+6114 
which has been previously 
associated \citep{hcg+01} with the EGRET source 3EG J2227+6122. The period of this pulsar 
is 52 ms, its distance is 
0.8 kpc \citep{boomerangdistance}, and the age is estimated to be 10.5 kyr 
and $\dot{\rm{E}}$ is $2.2 \times 10^{37}$ ergs/sec \citep{atnfcatalog,hcg+01}.  
Milagro detects a 6.6$\sigma$ excess at the position 
of the pulsar and a local maximum of 6.8$\sigma$. The peak of the 
Milagro excess is RA=22h28m17s Dec=60$^\circ$29m with a statistical position error of 0.36.
This source was reported as candidate C4 by Milagro in \citep{milagrogalacticplane}.
With the additional data and improved analysis presented here, this source is elevated 
to a high-confidence detection. Milagro also identifies this source as clearly not a point 
source, with a long extension to the south\footnote{Note added in press:  
The Fermi-LAT collaboration has submitted a paper announcing the discovery of
a new pulsar -- not included in the BSL -- with the current best position of RA=339.561, Dec=59.080 \citep{newpulsar_privcomm}.  
Milagro observes a 4.7$\sigma$ excess at the location of the pulsar.  It may be that the large size of the multi-TeV
emission associated with 0FGL J2229.0+6114 is in fact due to these two nearby sources.}


The remaining five objects with greater than 3$\sigma$ excess in the Milagro data have
not been previously detected above 1 TeV energies:


{\bf 0FGL J0631.8+1034} is the radio pulsar J0631+10
\citep{zcwl96}.  This pulsar has a period of 288 ms and an estimated 
age of 43.6 kyr, a distance of 6.55 kpc and $\dot{\rm{E}}$ of 
1.7$ \times 10^{35}$ erg/s \citep{atnfcatalog}. 
The VERITAS upper-limit for this region is
1.3\% of the Crab \citep{veritasgeminga}.

{\bf 0FGL J1844.1-0335} is unassociated with any known source.  It is
an interesting source because it occurs at a Declination at the
edge of Milagro's sensitivity and, if the Milagro observation is real, it is extremely
bright above 1 TeV.  It is in the region of the Galactic plane surveyed by HESS
\citep{hesssurvey} but was not detected.  To account for the HESS non-detection,
the source would have to be extended or 
have a very hard spectrum extending to high energy.

{\bf 0FGL J1900.0+0356} has no known associations.

{\bf 0FGL J1954.4+2838} is coincident with SNR G65.1+0.6
which has been associated with PSR 1957+2831 \citep{snrg65}.

{\bf 0FGL J1958.1+2848}  is a LAT-discovered pulsar that is
associated with the EGRET source 3EG J1958+2909 \citep{thirdegret}. 

{\bf 0FGL J2021.5+4026} is a LAT-discovered pulsar that is
coincident with the gamma-Cygni SNR.  This 
source is located in the Cygnus region that is detected by Milagro as having
a broad extended excess.

\begin{figure}
\plottwo{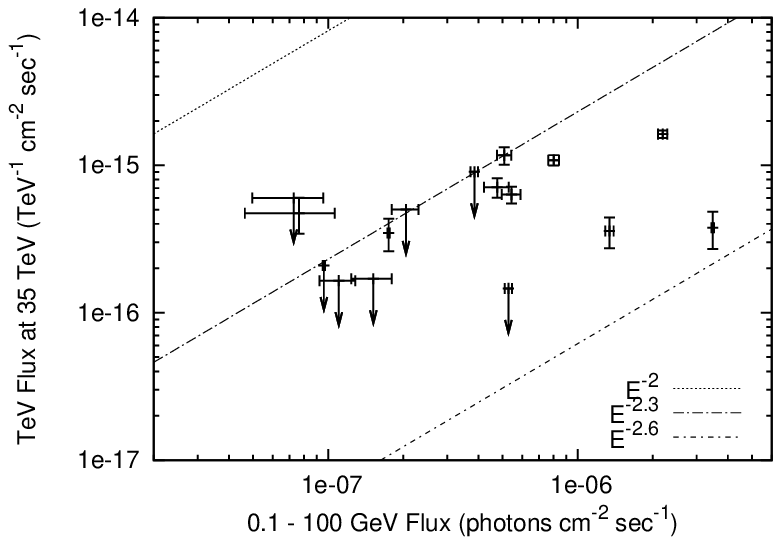}{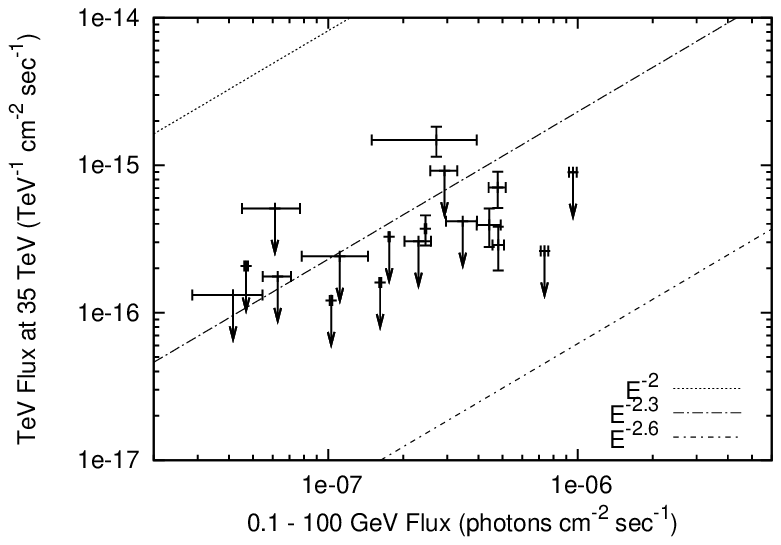}
\caption{\label{fluxes}
Flux estimates and upper limits for the 34 Fermi sources.  The horizontal 
axis quotes the integral Fermi flux from 100 MeV to 100 GeV and the 
vertical axis gives the Milagro flux or upper limit at 35 TeV.
Lines are 
shown with the extrapolation of the Fermi flux to Milagro energies, assuming
an $E^{-2.0}$, $E^{-2.3}$
and  $E^{-2.6}$ spectrum.  The left panel shows the results for 
the 16 pulsars and the right panel shows the results for the remaining 
18 sources.
}
\end{figure}

The relationship between the 
Fermi and Milagro source fluxes and 
upper limits for these 34 sources
is shown in 
Figure \ref{fluxes}.
The BSL values for the integral flux are shown with
the Milagro measurements of the differential flux at 35 TeV. 
The 35 TeV fluxes are roughly 
correlated with the measurements between 100 MeV and 100 GeV
but the correlation is not strong, 
with a correlation coefficient of the 3$\sigma$ points in log space
of only 0.2.  
One possible explanation for the pulsar
variation is that the pulsed emission is expected to be beamed (and thus 
viewing-angle 
dependent) and the unpulsed multi-TeV
emission is likely unbeamed \citep{gaenslerpwn}.  The spectrum that connects 
the Milagro flux to the
Fermi flux is universally softer 
than 2.0 and closer to 2.3, depending on the 
source.

We have found that the population of Fermi sources observed at or above 3$\sigma$ by
Milagro is dominated by pulsars and/or their associated PWN.  
Of the 4 high-confidence Milagro detections associated with pulsars of known periodicity
and distance, 3 (namely J0534.6+2201, J0634.0+1745, and J2229.0+6114) 
have $\dot{\rm{E}}/\rm{d}^2$ above $10^{35}$ $\rm{ergs}$ $\rm{s}^{-1}$ $\rm{kpc}^{-2}$
where $\dot{\rm{E}}$ is the spin down luminosity
and $\rm{d}$ is the distance to the pulsar.  The distance on the 
fourth (J2020.8+3649) is uncertain.  Using the 3-4 kpc distance
implied by x-ray measurements \citep{chandradragonfly} rather than the 
12 kpc measurement implied by the pulsar dispersion measurement, it too
has $\dot{\rm{E}}/\rm{d}^2$ above $10^{35}$ $\rm{ergs}$ $\rm{s}^{-1}$ $\rm{kpc}^{-2}$.
A similar association with high 
$\dot{\rm{E}}/\rm{d}^2$ pulsars is reported by HESS \citep{hesspulsars}.
Since the pulsed emission is beamed
and the PWN is not, all high $\dot{\rm{E}}/\rm{d}^2$ are possible candidates formulti-TeV
emission. 
We have
searched the ATNF pulsar database \citep{atnfcatalog} 
for northern-hemisphere pulsars 
with a high $\dot{\rm{E}}/\rm{d}^2$, which were not reported in the Fermi BSL.
Of the 25 highest $\dot{\rm{E}}/\rm{d}^2$ pulsars, 
there are 10 in the northern-hemisphere and 5
not identified as GeV sources by Fermi.
These 5 are J0205+6449, J0659+1414, J1930+1852,
J1913+1011 and J1740+1000. Of these, the largest statistical significance was 3.3 standard deviations
(PSR J1930+1852), not significant enough
to claim this as new source of multi-TeV
gamma rays (though follow-up observations 
are warranted).

Finally, it is interesting to note that of the 
sources published
in the Milagro survey of the Galactic plane\citep{milagrogalacticplane}, 
all 4 sources and two of the 4 source candidates are now strongly associated with pulsars, 
suggesting that most of the Milagro sources are multi-TeV
PWN.  
We also note the high 
efficiency with which MeV to GeV pulsars are observed above 1 TeV, and a 
qualitative picture is emerging where the typical Galactic multi-TeV source is a 
PWN associated with a MeV to GeV pulsar.



\acknowledgments

We gratefully acknowledge Scott Delay and Michael Schneider for their
dedicated efforts in the construction and maintenance of the Milagro
experiment. This work has been supported by the National Science Foundation
(under grants PHY-0245234, -0302000, -0400424, -0504201, -0601080,
and ATM-0002744), the US Department of Energy (Office of High-Energy Physics
and Office of Nuclear Physics), Los Alamos National Laboratory, the
University of California, and the Institute of Geophysics and Planetary
Physics.

\bibliography{bibliography}




\begin{deluxetable}{cc|cc|cc|cc|c}
\tabletypesize{\scriptsize}
\tablecaption{Summary of the search for TeV emission from sources in the 
Fermi BSL.  
\label{results}
}
\tablewidth{0pt}
\tablehead{
\colhead{} & \colhead{} & \colhead{} & \colhead{} 
& \colhead{} & \colhead{}& \colhead{Flux } & \colhead{} & \colhead{} \\
Name   & type & RA    & DEC   & $l$ & $b$ & ($\times$10$^{-17}$ TeV$^{-1}$  & Signif.      & TeV   \\
(0FGL) &      & (deg) & (deg) &(deg)&(deg)& sec$^{-1}$ cm$^{-2}$)           & ($\sigma$'s) & assoc.\\
}
\startdata

J0007.4+7303 & PSR &   1.85 & 73.06 & 119.69 &  10.47 & $<$ 90.4        & 2.6  &  \\
J0030.3+0450 & PSR &   7.60 &  4.85 & 113.11 & -57.62 & $<$ 20.9        & -1.7 &  \\
J0240.3+6113 & HXB &  40.09 & 61.23 & 135.66 &   1.07 & $<$ 26.2        & 0.7  & LSI +61 303 \\
J0357.5+3205 & PSR &  59.39 & 32.08 & 162.71 & -16.06 & $<$ 16.5        & -0.1 &  \\
J0534.6+2201 & PSR &  83.65 & 22.02 & 184.56 &  -5.76 & 162.6 $\pm$ 9.4 & 17.2 & Crab \\
J0613.9-0202 & PSR &  93.48 & -2.05 & 210.47 &  -9.27 & $<$ 60.0        & -0.0 &  \\
J0617.4+2234 & SNR$^a$ &  94.36 & 22.57 & 189.08 &   3.07 & 28.8 $\pm$ 9.5  & 3.0  & IC443 \\
J0631.8+1034 & PSR &  97.95 & 10.57 & 201.30 &   0.51 & 47.2 $\pm$ 12.9 & 3.7  &  \\ 
J0633.5+0634 & PSR &  98.39 &  6.58 & 205.04 &  -0.96 & $<$ 50.2        & 1.4  &  \\
J0634.0+1745 & PSR &  98.50 & 17.76 & 195.16 &   4.29 & 37.7 $\pm$ 10.7 & 3.5  & MGRO C3 \\
             &     &        &       &        &        &                 &      & Geminga\\
J0643.2+0858 &     & 100.82 &  8.98 & 204.01 &   2.29 & $<$ 30.5        & 0.3  &  \\ 
J1653.4-0200 &     & 253.35 & -2.01 &  16.55 &  24.96 & $<$ 51.0        & -0.5 &  \\
J1830.3+0617 &     & 277.58 &  6.29 &  36.16 &   7.54 & $<$ 32.8        & 0.2  &  \\ 
J1836.2+5924 & PSR & 279.06 & 59.41 &  88.86 &  25.00 & $<$ 14.6        & -0.9 &  \\
J1844.1-0335 &     & 281.04 & -3.59 &  28.91 &  -0.02 & 148.4 $\pm$ 34.2& 4.3  &  \\
J1848.6-0138 &     & 282.16 & -1.64 &  31.15 &  -0.12 & $<$ 91.7        & 1.7  &  \\
J1855.9+0126 & SNR$^a$ & 283.99 &  1.44 &  34.72 &  -0.35 & $<$ 89.5        & 2.2  &  \\
J1900.0+0356 &     & 285.01 &  3.95 &  37.42 &  -0.11 & 70.7 $\pm$ 19.5 & 3.6  &  \\
J1907.5+0602 & PSR & 286.89 &  6.03 &  40.14 &  -0.82 & 116.7 $\pm$ 15.8& 7.4  & MGRO J1908+06 \\
             &     &        &       &        &        &                 &      & HESS J1908+063 \\
J1911.0+0905 & SNR$^a$ & 287.76 &  9.09 &  43.25 &  -0.18 & $<$ 41.7        & 1.5  &  \\ 
J1923.0+1411 & SNR$^a$ & 290.77 & 14.19 &  49.13 &  -0.40 & 39.4 $\pm$ 11.5 & 3.4  & HESS J1923+141 \\ 
J1953.2+3249 & PSR & 298.32 & 32.82 &  68.75 &   2.73 & $<$ 17.0        & 0.0  &  \\ 
J1954.4+2838 & SNR$^a$ & 298.61 & 28.65 &  65.30 &   0.38 & 37.1 $\pm$ 8.6  & 4.3  &  \\ 
J1958.1+2848 & PSR & 299.53 & 28.80 &  65.85 &  -0.23 & 34.7 $\pm$ 8.6  & 4.0  &  \\ 
J2001.0+4352 &     & 300.27 & 43.87 &  79.05 &   7.12 & $<$ 12.1        & -0.9 &  \\ 
J2020.8+3649 & PSR & 305.22 & 36.83 &  75.18 &   0.13 & 108.3 $\pm$ 8.7 & 12.4 & MGRO J2019+37 \\
J2021.5+4026 & PSR & 305.40 & 40.44 &  78.23 &   2.07 & 35.8 $\pm$ 8.5  & 4.2  &  \\
J2027.5+3334 &     & 306.88 & 33.57 &  73.30 &  -2.85 & $<$ 16.0        & -0.2 &  \\ 
J2032.2+4122 & PSR & 308.06 & 41.38 &  80.16 &   0.98 & 63.3 $\pm$ 8.3  & 7.6  & TEV 2032+41   \\
             &     &        &       &        &        &                 &      & MGRO J2031+41 \\
J2055.5+2540 &     & 313.89 & 25.67 &  70.66 & -12.47 & $<$ 17.6        & -0.0 &  \\ 
J2110.8+4608 &     & 317.70 & 46.14 &  88.26 &  -1.35 & $<$ 24.1        & 1.1  &  \\ 
J2214.8+3002 &     & 333.70 & 30.05 &  86.91 & -21.66 & $<$ 20.7        & 0.6  &  \\ 
J2229.0+6114 & PSR & 337.26 & 61.24 & 106.64 &   2.96 & 70.9 $\pm$ 10.8 & 6.6  & MGRO C4 \\
J2302.9+4443 &     & 345.75 & 44.72 & 103.44 & -14.00 & $<$ 13.2        & -0.6 &  \\

\enddata
\tablecomments{
The source identity in the 0FGL catalog is given
with the source location in celestial and galactic coordinates.  
We give the measured flux for all sources above 3$\sigma$ at a characteristic 
median energy of 35 TeV.  The 2$\sigma$ upper limits are given for other sources.  
The statistical significance and nearby TeV associations are noted.  
}
\tablenotetext{a}{The BSL association with a known SNR is based on 
similar location.}

\end{deluxetable}

\newpage

\section*{Erratum: Milagro Observations of Multi-TeV Emission from Galactic Sources in the Fermi Bright Source List (2009, ApJ, 700L, 127A)}

The position of the peak of the Milagro excess coincident with {\bf 0FGL J2229.0+6114}
was incorrectly reported as RA=22h28m17s Dec=60$^\circ$29m. The correct 
position is RA=22h28m44s Dec=61$^\circ$10m with a statistical
position error of 0.165$^\circ$.

\end{document}